Opinion paper

# Open microscopy in the life sciences: Quo Vadis?

Johannes Hohlbein[1,2,*], Benedict Diederich[3,4], Barbora Marsikova[3], Emmanuel G. Reynaud[5], Séamus Holden[6], Wiebke Jahr[7], Robert Haase[8], and Kirti Prakash[9]

[1] Laboratory of Biophysics, Wageningen University & Research, Wageningen, the Netherlands

[2] Microspectroscopy Research Facility, Wageningen University & Research, Wageningen, the Netherlands

[3] Leibniz Institute for Photonic Technology, Jena, Germany

[4] Institute for Physical Chemistry, Friedrich-Schiller University, Jena, Germany

[5] School of Biomolecular and Biomedical Sciences, University College Dublin, Dublin, Ireland

[6] Centre for Bacterial Cell Biology, Biosciences Institute, Newcastle University, Newcastle upon Tyne, UK

[7] In-Vision Digital Imaging Optics GmbH, Austria

[8] DFG Cluster of Excellence "Physics of Life", TU Dresden, Germany

[9] National Physical Laboratory, Teddington, UK

Contacts:

johannes.hohlbein@wur.nl (https://orcid.org/0000-0001-7436-2221),

benedictdied@gmail.com (https://orcid.org/0000-0003-0453-6286),

marsikova.b@gmail.com (https://orcid.org/0000-0002-8206-1549),

emmanuel.reynaud@ucd.ie (https://orcid.org/0000-0003-1502-661X),

Seamus.Holden@newcastle.ac.uk (https://orcid.org/0000-0002-7169-907X),

wiebke.jahr@web.de (https://orcid.org/0000-0003-0201-2315),

robert.haase@tu-dresden.de (https://orcid.org/0000-0001-5949-2327),

kirtiprakash2.71@gmail.com (https://orcid.org/0000-0002-0325-9988)

[*]corresponding author

keywords: open microscopy, open source, open science, open hardware

**Abstract**

Light microscopy allows observing cellular features and objects with sub-micrometer resolution. As such, light microscopy has been playing a fundamental role in the life sciences for more than a hundred years. Fueled by the availability of mass-produced electronics and hardware, publicly shared documentation and building instructions, open-source software, wide access to rapid prototyping and 3D printing, and the enthusiasm of contributors and users involved, the concept of open microscopy has been gaining incredible momentum, bringing new sophisticated tools to an expanding user base. Here, we will first discuss the ideas behind open science and open microscopy before highlighting recent projects and developments in open microscopy. We argue that the availability of well-designed open hardware and software solutions targeting broad user groups or even non-experts, will increasingly be relevant to cope with the increasing complexity of cutting-edge imaging technologies. We will then extensively discuss the current and future challenges of open microscopy.

**A. Introduction**

Light microscopy has been pivotal in the life sciences to study small features and objects otherwise hidden to the naked eye. Simple microscopes such as the Foldscope or smartphone auxiliary lenses are available for a few Euros and form the basis of citizen science projects, scientific education, and medical diagnosis[1,2].

Driven by societal and academic shifts towards a culture of open science and open technology, more and more information on modalities for advanced and high-end microscopy has become openly available to interested researchers and private enthusiasts. Compared to the open software movement, however, the development of open hardware frameworks for microscopy accelerated only rather recently with the increasing accessibility and affordability of suitable components. With 3D printers, rapid reproduction of designs and prototyping moved from professional machine shops to the living room table. Web shops deliver sophisticated CNC milled designs within days to the doorsteps. Mass-produced electronics such as light-emitting or laser diodes, microcontrollers, optical parts, and (mobile phone) cameras have significantly levelled the playing field for high-end instrumentation. Open microscopy does not automatically imply that solutions are cheap. Some of the frameworks further discussed below feature higher-priced components such as microscope objectives, laser engines, and cameras often costing tens of thousands of Euros. As such, these components maximize performance but can surprisingly often be adequately replaced by cheaper alternatives that better benefit from the economies of scale[9].

Supported by their accessible documentation, all open projects empower scientists to adopt solutions - even if only as a source of inspiration. In academia, open hardware and software projects are lifesavers at a time of fierce competition for funding and a decline in government investment in science. Projects that have developed strong communities enable researchers to receive support within minutes in public support forums and over social media.

By its very nature, microscopy is interdisciplinary combining aspects from physics, engineering, informatics, biology, and data science, often with a hint of breakthrough technologies. As educational resources, open microscopy frameworks further provide an entry to the world of coding, tinkering and science with minimal thresholds to overcome.

In an age of seemingly unlimited educational and technical possibilities, what does that mean to the general field of optical light microscopy and its high-end applications? What are the remaining challenges and where lie potential issues? Here, after providing a brief overview of current open-hardware and open-software microscopy projects, we highlight several goals and challenges that we consider important to ensure the growth and future success of open microscopy.

## B. On the purpose of open science and open microscopy

Open science, open source, open access, open hardware, open *everything* is now the motto for many communities and has become the obligatory reference. In general, open science seeks to improve transparency, reproducibility, inclusiveness, and accessibility of research and innovation as, for example, discussed in the UNESCO draft recommendation on open science[10]. For scientific research, the conventional setting of an academic environment kept the science largely behind closed doors; although results were published, information on the methodology, the experimental settings or access to raw data was, and unfortunately often still is, hard to come by (Figure 1). Open science is addressing these issues by providing a myriad of additional interaction points between researchers and more broadly between citizens, starting from the collection of ideas over reusing hardware designs and open-source software to the exploration of publicly available data by anyone interested. For scientific data, the FAIR principle (Findable, Accessible, Interoperable, Reusable)[11] provides guidelines for moving towards what science should be all about: shared knowledge accessible to all. With that in mind, we here define open microscopy as scientific data which contains information on (i) how to build, maintain and use (light) microscopes, (ii) how to prepare, handle and measure samples and (iii) how to analyze, store and distribute experimental data and models.

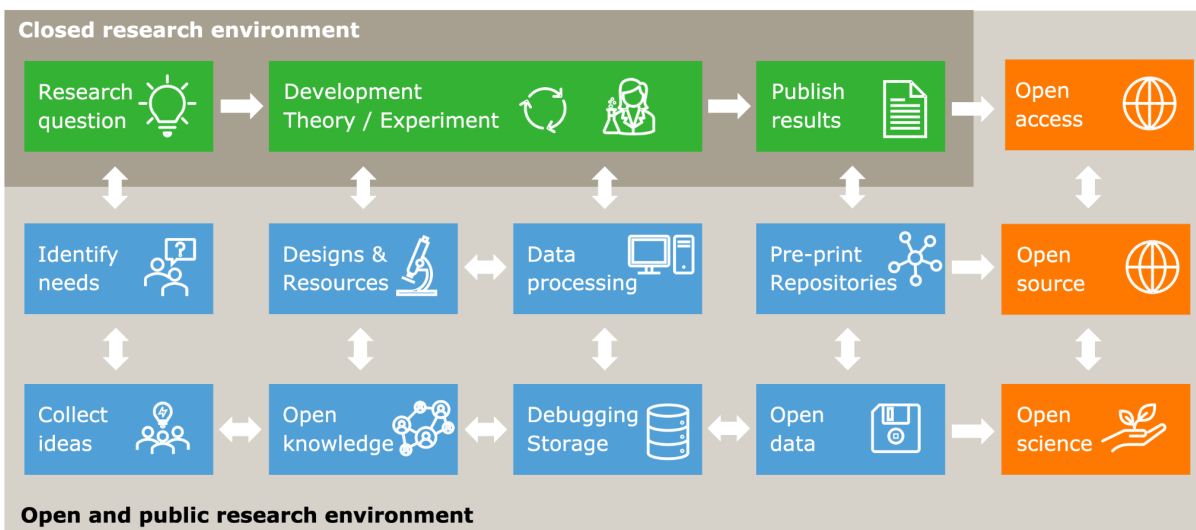

**Figure 1.** Open science can accelerate scientific progress. Instead of operating in a closed environment, representing, for example, individual labs, research units or scientific areas, free flow of information and exchange of ideas, resources and data enables the efficient pooling of resources.

### *B.1. Open microscopy is just good scientific practice*

For open hardware, recent work highlights general opportunities and best practices[12,13]. For light microscopy, this implies making new imaging technologies accessible and easily reproducible. In the past, new methods were often published with only a vague cartoon of the optical setup and without access to any software or electrical circuit diagrams. Even for specialist labs, implementation of published modalities often required extensive reverse engineering and tinkering for many years. Commercial microscopes, on the other hand, are often built on closed-source methods and compound the reproducibility problem by making key steps in the process of biological discovery dependent on an ill-characterized black box. Open microscopy can address this problem by ensuring that any new imaging method, both hardware and software, is sufficiently documented and open to allow straightforward reproduction by other specialists, and without information being hidden behind restrictive Material Transfer Agreements (MTA) or Non-Disclosure Agreements (NDA). Documenting assembly and manufacturing (e.g., 3D printing, CNC machining) guides, bill of materials or even providing video tutorials that help others recreating a setup without additional help by the authors, acts as an amplifier, in which knowledge is multiplied and can be used at remote places outside the originating lab. It can be very motivating to see one's work being replicated and used by others, even if current scientific scoring systems may only recognize formal citations and not pre-publication replications.

Since prototype microscopes usually require extensive further development to make them production-ready and user-friendly, this does not preclude their co-existence with commercial devices. The need for users to receive comprehensive product support and a turnkey solution is not necessarily a requirement of the Open Microscopy movement. Generally, we argue that open microscopy should be considered an essential requirement for any scientific publication of a new microscopy method to meet modern standards of scientific reproducibility. Detailed documentation, including why something was implemented in the suggested way, is also key for others to learn about the given technique and to discover potential optimization steps. For small hardware or software components, this could further imply making drawings or source code available under an open-source license. In general, documentation can be written in the form of citable scientific publications and thus, is well compatible with the academic incentive system. Editors from the optics and scientific imaging fields, however, should increase efforts in supporting scientists, who are willing to share their designs openly, to improve the reproducibility of publications and science in general.

Reproducibility in the context of data analysis implies that both the experimental or simulated data as well as the analysis software is openly available. We therefore argue that reproducibility is the key to advancing science as without rigorous verification of results and discoveries true scientific progress is impossible.

Academic researchers should be aware that, by default, everything developed and created is the property of the research institute, meaning that researchers leaving the institute may lose both rights and access to their unpublished intellectual contributions. To permit the use and further development of open microscopy projects by anyone, regardless of location or affiliation, appropriate licenses such as CERN Open Hardware License, MIT, GPL v3 or Creative Commons must be chosen. This also addresses the issue posed by active patents that theoretically prohibit the replication and productive use of protected setups in the laboratory as discussed here[14]. We recommend scientists and developers to make themselves aware of the regulations and possibilities with the institutional IP handling offices as early as possible.

While not an intrinsic feature of open source, we encourage developers to use version control tools like git (GitHub, Gitlab) at any stage of the project to document the process and track individual contributions that can help to clarify the contribution of individual authors.

***B.2. Open microscopy enables flexible and powerful platforms for different user groups***

Until recently, microscopy hardware developers seeking to develop new optical methods faced the choice of either retrofitting new hardware onto an existing commercial microscope (body) or designing and building an entire bespoke microscope from individual optical components. Neither fits the requirements of a hardware developer perfectly. Although using a commercial body has the advantage of starting with a working system that usually lacks serious optical aberrations and offers a stable mechanical base, critical optical planes required in the microscope development process are often hidden, unknown or inaccessible within the monolithic body. Furthermore, some features implemented for user-friendliness and safety (eyepieces, safety interlocks, dedicated software) can potentially prevent the hardware developer from easily modifying a setup. Customized designs, on the other hand, offer full control and often lower costs at the expense of significant development time and the need to re-implement basic components and features, such as focusing or sample positioning. One drawback of custom microscopy solutions is that they are often less user-friendly compared to their commercial counterparts that offer safe housing or one-click software solutions. In the end, commercial bodies and self-designed systems may target different audiences but, as we will discuss in the following, the gap has been closing rapidly.

Rather than classifying open hardware frameworks in terms of their apparent complexity, we suggest looking at potential user groups. We foresee disruptive potential of extremely simplified, minimalist open microscopes (FlyPi[15], OpenFlexure[3], UC2 system[4], µCube[16], Octopi[17]) as they change advanced microscopy from a scarce resource shared between many labs to everyday tools of life scientists and hobby enthusiasts all over the world (Figure 2). These microscopes are designed to be affordable, adaptable, reproducible, and exchangeable or easily repairable, for example using 3D printed parts instead of specialist components as recently reviewed for light microscopy in reference[18]. Time-intensive experiments such as live cell microscopy or automated high-content screening could be made easily accessible, creating new scientific opportunities similar to how low-cost (whole-genome) DNA sequencing enabled the widespread use of genomics in biological research.

For developmental biologists interested in volumetric imaging, the OpenSPIM (SPIM: Selective- Plane Illumination Microscopy) project has been a successful example of an open microscopy platform that required an entire redesign of microscopy hardware and software[19]. Starting in 2010, OpenSPIM enabled many labs as well as imaging platforms to build and apply light-sheet microscopy at a time when commercial solutions were neither accessible nor affordable.  It is estimated that more than 80 systems have been built with at least 6 in facility settings. Similarly, the mesoSPIM initiative targets the imaging of large samples and provides a very comprehensive open-source documentation[20]. According to the project website (https://mesospim.org/), at least 12 builds are in operation and follow-up work provides detailed protocols for tissue clearing[21]. Further, an open platform for scanned oblique plane illumination microscopy (SOPi), was introduced that features open hardware assembly, clear alignment protocol, and control software for sub-micron resolution single objective light-sheet microscopy[22].

A number of open microscopy frameworks have been developed that more closely resemble the layout of conventional upright commercial systems but offer a higher degree of modularity and customizability. Several platforms for cost-effective single-molecule localization microscopy visualize biological structures with a resolution of well below 50 nm (WOSM[23], liteTIRF[24]) or even characterize protein-DNA-RNA interactions in live bacteria (miCube[25]). Additional frameworks focus on conventional epifluorescence microscope (LFSM[26]), high-throughput screening and tracking of microorganisms (Squid[27], cell biology[28]), diffusion-based confocal for analyzing conformational dynamics of biomolecules[29] or detection of protein aggregation[30], two-photon $Ca^{2+}$ deep tissue imaging[31], and structured illumination for sub-diffraction resolved (live) cell imaging[32,33] (Figure 2). Depending on exact implementations, the cost of these microscopes can be considerably lower than for commercial systems, although, as discussed below, the costs due to expert time investment for both build and maintenance should not be underestimated.

We note that every published technical solution can potentially contribute to simplifying the development of future microscopy techniques. Open microscopy software is also constantly evolving. ImageJ[6] and Fiji[7] have been pivotal for analyzing images and movies by life scientists for many years. Their microscope control sibling µManager[35] and counterparts in the Python world[36–39] have been changing the way how the community analyses and displays images in the microscopy context for many years, with new plugins and functionalities for data acquisition being constantly released[40,41]. Smooth transfer of images acquired on one system to another was key to create this ecosystem for bioimage analysis that now flourishes beyond the manufacturer formats, especially due to the efforts of the Open Microscopy Environment (OME) initiative[42]. The field of image data analysis has exploded in recent years now involving deep learning techniques for image quality improvements, segmentation and overall data analysis (e.g., CARE[48], StarDist[49], CellPose[50], QuPath[51], ZeroCostDL4Mic[8]). Modern Graphics Processing Units (GPUs) do not just execute deep learning algorithms in the shortest amount of time, the technology also supports post-processing of data in real-time such as multiview deconvolution[52], deconvolution and fusion[53], and general-purpose image processing[54].

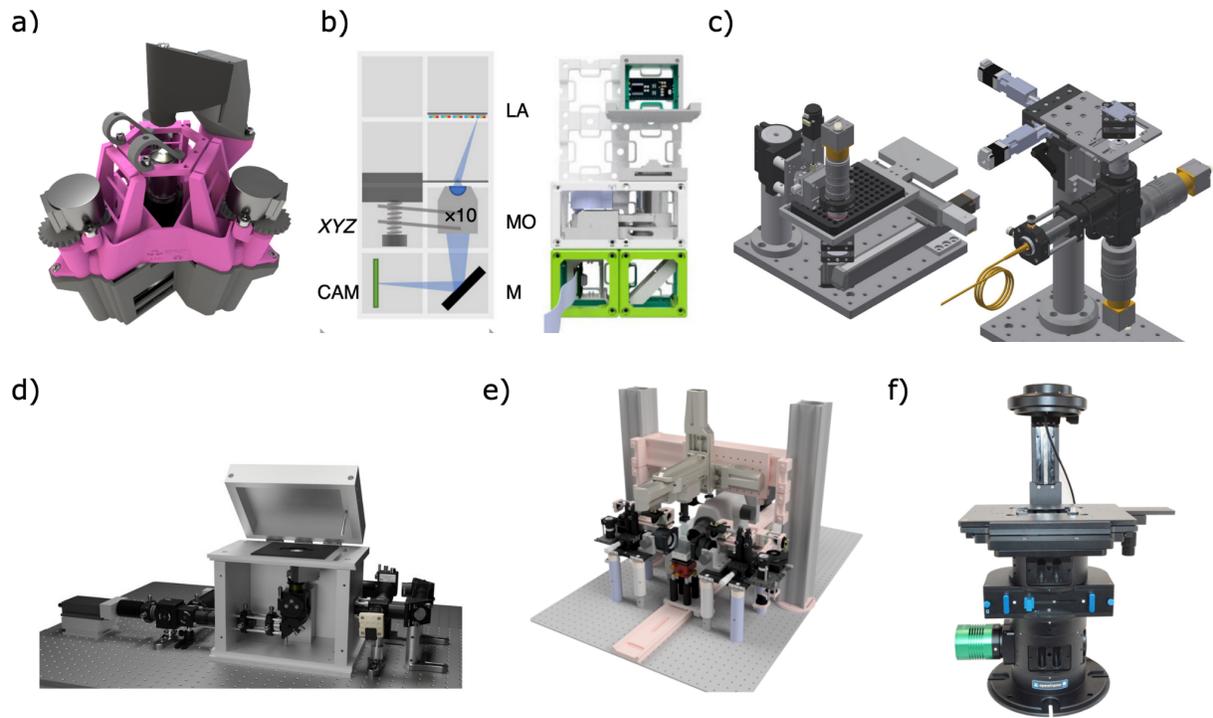

**Figure 2.** Overview on various open-microscopy projects. **a)** OpenFlexure[3], **b)** "You see, too" (UC2[4]), **c)** Simplifying Quantitative Imaging Platform Development and Deployment (Squid[27]), **d)** single-molecule Förster Resonance Energy Transfer box (smFRETbox[29]), **e)** MesoSPIM[20], **f)** openFrame[34], an open microcopy framework that is commercially available. Permissions: images a), b), d) reproduced under CC-BY 4.0, c), e) under CC-BY-NC-ND 4.0 and f) with permission from Cairn-Research).

## C. Current and future challenges

### *Open microscopy: diverse but united*

Although involved developers have a broad set of goals and applications, a common conviction can be found: the principles of open science have the power to change microscopy for the better, leading to more innovation, accessibility, and reproducible science. An implementation of open microscopy in its purest form in which ideas and experimental designs are shared online and on social media sometimes even before experimental realizations take place has been pioneered by Andrew York (https://andrewgyork.github.io/). In the following, we discuss some of the key challenges.

### *C.1. Accessibility, availability, safety, and time versus money*

The reasons for working on and with open microscopy projects are as diverse as the people involved. Some might enjoy the tinkering aspects most (the tinkerer). Others see open software and technology primarily as tools to address their scientific questions as affordable and fitting as possible (the end-user). Microscope developers and end-users, and all the researchers falling somewhere in between, may have a very different vision for open microscopy and should be aware of each other. The end-user is likely to prefer more polished software or hardware, sometimes even willing to sacrifice additional features for stability and ease of use. In fact, the end-user might have less time to build or adapt complete solutions and would rather prefer to buy them. Both sides ultimately depend on each other like in a classical "supply and demand" situation in which a growing request for innovative solutions can support people working on them. Along the same line, many projects require substantial training and build-up of expertise before productive data acquisition can be achieved. Although proper documentation and design can support the perspective of the end-user, the availability of personalized training and education is needed. As we will discuss later, education is an inherent feature of open microscopy since parts to build microscopes can easily be sourced.

One often-used statement in the open microscopy field is that an instrument was designed for (part-) costs that are ten times cheaper than the price of a comparable commercially available instrument. We consider such statements at best misleading and at worst plainly wrong as neither the costs of development nor the time spent to build the instrument is accounted for in most cases. We also point out that any company must fulfil a minimum of conformity with health, safety, and environmental protection standards (e.g., CE, FCC, TÜV or others) for their products and provide customer support. In open projects, even when building up with commercially available components, the sole responsibility on safety is shifted to the user. Additionally, customer support depends largely on the goodwill of the developer.

### *C.2. Standards and continuing proliferation: it's not a bug, it's a feature!*

As outlined above, the number of open microscopy frameworks has hugely increased over the last years. An interesting example of proliferation is the variety of software packages available for single-molecule localization microscopy that fit data from individual fluorescent emitters to obtain their position with sub-pixel accuracy. Recently, over 30 different software packages were compared using a diverse set of metrics[43]. Although such wide choice can be intimidating at first, the robust assessment itself, indicating strengths and weaknesses of individual software packages, is excellent proof of the benefits of open microscopy. As new packages and developments can be directly compared to previous ones, the super-resolution field was able to quickly reach a high degree of maturity in terms of data analysis. It became equally apparent that user-friendly implementations of the best approaches are critical to reaching end-users.

Decentralized hardware development in which experts design solutions for specific problems cannot often be integrated in other frameworks. Here, interfaces such as device adaptors in µManager[35] or mechanical adaptors as provided by the modular optical toolbox UC2[4] can help to bridge multiple different hardware and software projects. Recently, a series of documents suggest the implementation of standards in open hardware and software development[44] as well as data provenance and quality control in microscopy[45,46]. We suggest that these best practices are requested and followed by scientists, reviewers, and editors to ensure and enable long-lasting inter-device operability. We note, however, that hardware- and software development significantly differ in the way how solutions are distributed. Ready-to-install software packages require only a limited knowledge about the underlying system integrity. Well-formulated building instructions still require someone who gathers all necessary parts (e.g., 3D printed or from external sources), assembles these parts and invests a fair amount of time to finally get the system running - no matter how well the documentation is described. The equivalent to a ready-to-install software package could be assorted hardware kits (UC2[4]), but the non-zero cost of duplicating hardware poses a challenge for widespread distribution. It remains important to implement and use calibration control points and sample standards both for hardware as well as software.

### C.3. Generating shareable hardware files

Open microscopy projects are shared using different files and data formats. Whereas many formats are open and suitable viewers are freely available, this is not necessarily the case for designs that feature Computer Aided Design (CAD). For CAD, buying suitable software licenses can be prohibitively expensive leading to limited accessibility and unknown design history. For 3D printing, 3D models can be exported, for example, in the *.stl format which describes only the surface geometry of a three-dimensional object without any scale thereby inhibiting any modifications on the design. Alternatives, such as sharing links to cloud-based CAD software (e.g., Fusion360, Tinkercad) or relying on open-source CAD models (e.g., openSCAD, Python-based FreeCAD) can help to distribute design files across many different development environments. Ultimately, publishers and developers should ensure that design files are available as (vectorized) raw data-file as well as processed files. The open source hardware association (OSHWA) offers a guideline to maintain a good quality of openness in projects (https://www.oshwa.org/sharing-best-practices/). Whilst protein structures, DNA sequences and other bioinformatic data can be deposited in well-established databases (e.g., PDB, EBI), no dedicated repositories for microscopy hardware are currently available.

### C.4. Connecting open-source software to open hardware

The close connection between open hardware and software is inevitable for complex microscopy projects. Projects such as µManager[35], Pycro-Manager[36] and Python-microscopy[37,47] have been playing a key part in connecting setup control, data acquisition and data analysis. When it comes to hardware control, the availability of open-source device drivers and adapters is crucial. The software architecture used in µManager, for example, standardizes how hardware devices can be controlled from diverse software components via a plug-in mechanism. If this mechanism is well-designed and easy to use, developers from all over the world will contribute plugins. As a case in point, the µManager community managed to collect hundreds of device adapters (https://micro-manager.org/Device_Support).

Many of the earlier introduced approaches for computational microscopy have the potential to lower the quality requirements for microscopy hardware. Efficient denoising algorithms, for example, allow the use of lower laser powers for fluorescence excitation or the use of cheaper cameras without compromising final image quality.

Such advanced analysis techniques, however, have not been integrated widely and reliably into any open microscope control software yet. Combining open software solutions for microscope control, image processing and data analysis is hugely challenging, requiring many developers from different backgrounds closely working together to optimize signal and data streams. First promising steps have been made by the smart microscopy software autopilot[55] and by combining OpenFlexure, ImJoy and UC2[56]. These scenarios highlight how important interdisciplinary collaboration is when it comes to solving problems in biology using experimental imaging and software approaches. Overall, plugin-based software projects that provide a reusable development platform can help integrating, e.g., graphical user interfaces (see Fiji, napari, ImJoy). Plugins can be easily shared with the community; algorithms and code can thus be used without much prior knowledge, which leads to a significantly increased acceptance by users. Open-source licenses enable the free (re-)use of software and thus contribute to accelerated research, as solutions are not always created from scratch.

### *C.5. Strategies enabling long term support of open-microscopy projects*

From our experience, many microscopy projects are initially driven by one or two people and have a limited time as scientific advancements can render ideas and concepts obsolete very quickly. Some projects, however, develop into large community-driven projects with specific requirements to gain and maintain long-lasting relevance and impact.

When it comes to questions regarding long-term support, clear communication between the developer and the potential target audience is advised to keep expectations aligned and in check. Developers should indicate as soon as possible whether their project is intended as a research platform for others that could turn into a community-driven project or whether the developer is mainly interested in using their hardware or software to promote their research. Gathering a broad community can push projects forward and active communities often generate new ideas that developers can incorporate. A good example is the OpenFlexure community, which has a Helpdesk/Forum where users can exchange problems, present results, and share ideas (https://openflexure.discourse.group/). For the primary developer, providing this kind of service plus managing the contributions of others comes at substantial costs, which are often difficult to cover in the current academic incentive system and can put a strain especially on smaller groups. Although funding agencies widely propagate the idea of open science, institutional support or open calls that are explicitly dedicated to the development of open hardware, software, and knowledge exchange projects are rare. The Chan Zuckerberg Initiative and NASA are notable exceptions providing substantial funding to support open science.

Furthermore, interacting with the community, selecting issues to work on and motivating others to support an open hardware initiative requires a substantial investment of time and effort. We recommend developers who want to develop community standards to think about these aspects carefully and identify sources and people for support early, noting that follow-up costs, both in time and money, cannot be paid by a single PhD student or postdoc no matter how enthusiastic they are.

What are the requirements for microscopy projects to succeed as community standards? **First** and foremost, we think that a clear need should be identified by the developer/community. As a recent example, the need for an adaptable image viewer interacting with the Python environment kickstarted the impressive and community-driven development of napari (https://zenodo.org/record/3555620). **Second**, some *uniqueness* to the approach differentiating it sufficiently from existing solutions should exist. If uniqueness is not sufficiently given, we recommend contributing to the other project. **Third**, one or more core developers with sufficient resources in terms of time, money, or appreciation are required to ensure continuity. **Fourth**, create and maintain an active user based on all levels of involvement ranging from "use as is", "test and report bugs", "request features" to "fix bugs and implement small features". The "miniscope"-community offers a great example how the base assembly, a miniaturized fluorescence microscope that enables *in vivo* brain imaging of mice, is used by the community to add a variety of soft- and hardware addons, such as two-photon imaging of one-shot 3D imaging using lightfields further reviewed in reference[57]. We note, however, that "open source" should not be translated as "free support". People already involved in a project have to talk to other potential users, help them, create examples, code/develop for them, work in close contact with them, find out what they want, create a product what they are willing to use without extra effort, give them the feeling of "YOU DID IT!". **Fifth**, merging expertise by means of adapting hardware or software designs from different projects, such as taking advantage of the microscopy OpenFlexure stage within the UC2 system can speed up development processes. **Sixth**, developers should strive for device interoperability by means of openly developed interfaces. Here, the software community gives a nice example of how object-oriented and packaged code enables simple reuse and hence speeds up development. **Seventh,** extensive high-quality documentation should be seen as a core priority. This documentation allows new users and developers to easily join and continue a project even if initial contributors left or initial investments have run dry.

We note that larger imaging facilities are well suited to support developers and users. We hope that universities and funders recognize the potential value of having a wide portfolio of maintained open microscopy projects.

### *C.6. Commercialization of open-source projects: Why and how?*

We think that it can be advantageous especially for hardware projects if (parts of the) assemblies could be made commercially available. We see an increasing demand for affordable and proven high-quality microscopy solutions by end-users who are not interested in building scientific instrumentation. In the simplest case, 3D printed or CNC-milled entities (e.g., OpenFlexure or miCube) can be sold directly, or prototypes can be made available that still require further adjustments to function to the user's need in the form of do-it-yourself kits. In other cases, microscopy solutions could become refined and user-ready products. For this route, however, there are many challenges to overcome. **First**, investors required to finance the transition from a prototype to a full product generally prefer solutions that are patent protected or protectable. **Second**, within universities, huge overhead costs often make the exploration of commercialization expensive and time-consuming. **Third**, the size of the market might be too small to get sufficient return on investment to keep the business viable. **Fourth**, there is the risk that potential patent infringement is targeted aggressively by established companies as soon as patented technology leaves the realm of pure academic use. **Fifth**, there is often a lack of knowledge of academics in the areas of economics and experience of how a project is turned from a prototype into a product. **Sixth**, the reluctance of some academics to devote part of their time to setting up a business.

One recent example of open microscopy hitting the shelves is the openFrame microscope developed by the French group and commercially available via Cairn research (https://www.cairn-research.co.uk/product/openframe-microscope/). Other examples of successful companies that rely on open hardware are Opentrons or Prusa 3D printers, which show that open source and economic efficiency are not mutually exclusive.

We hope that universities and their technology transfer units can develop solutions that reach the market with minimal bureaucratic and financial overhead for involved researchers. One potential route is involving external companies specializing in the commercialization of academic ideas and products. Some business models and companies even permit the production and sale of open-source hardware under open-source hardware licenses such as the CERN Open Hardware License (OHL). For a discussion on potential business models, the reader is referred to Josuah Pearce's essay[58].

When thinking about routes towards commercialization, another business opportunity could be to provide services related to specific open microscopy projects. Scientists who prefer to work with open solutions may neither have the experience nor the time to do these modifications and extensions themselves. Inviting a specialised and independent open hardware or software developer as a guest scientist or consultant might be more effective than hiring a postdoc. Such a job profile, however, still needs to be established and supported by research institutions.

In general, the field strongly requires role models; people that go from open source to commercialization and talk about it. Conferences, as well as journals, should invite people to talk and write about these important topics showing that (open source) business models can be sustainable.

*C.7. Continuing training and education*

The increasing complexity of methods and tools required for research in the life sciences requires continuing training and education. The financial investment necessary for hands-on training in optics and related fields has been substantially reduced with open instrumentation and simplified hardware demonstrating basic physical principles while mainly preserving quality in terms of resolution, usability, and stability. Moreover, in the interdisciplinary area of microscopy, project-based courses encourage creativity and the development of new approaches to solving individual user problems. Open education in microscopy further improves hardware projects via bidirectional exchange of knowledge and experience.

With the widespread use of digital teaching and learning platforms, and the possibility of printing or building the microscope yourself or converting a smartphone into one, training no longer has to take place at one location. Like in the flipped classroom concept, the tasks are discussed first, possibly online, solved individually outside the classroom and the results are discussed afterwards. During the SARS-CoV-2 pandemic, for example, the possibility of distributing UC2 boxes offered a hands-on practical course at times at which in-person lectures and lab work were not possible.

Low-cost microscopes like the Foldscope and the associated worldwide dissemination of small microscopes in places where they would otherwise not be found can lead to discoveries that can be shared and discussed both in class and with the wider community, e.g., on social networks. The associated ease of access to these tools, which are available to virtually everyone, makes education more inclusive. These novel ways of education also support the growing interest in STEM subjects. Training young interdisciplinary professionals with the help of open-source tools promote and create international cooperation. An important element for the future is making the resources comprehensive to reduce the burden on the educators and provide the easiest possible access for direct use in the classroom.

## D. Conclusion

In the past, advanced optical light microscopy was seen as an expensive specialist endeavor. Open microscopy, similar to the open-source movement as a whole, breaks down barriers in microscopy at all levels. We are convinced that this applies just as much to cutting-edge microscopy-driven research as it does to applications of simpler microscopy methods in areas such as healthcare and education.

Method developers working to expand the absolute technical limits of the field benefit from open microscopy as they build upon the latest, most sophisticated system designs. Instead of expending limited research time reproducing ill-documented systems, they can focus on the genuine novelty in their project. Thus, the complete and detailed documentation inherent to open microscopy drives the development of more sophisticated, more powerful new microscope technologies. Any new microscope development project will strongly benefit from the availability and accessibility of smart and open solutions for hardware, software, and assays. In fact, we do not expect any future cutting edge microscope development to be possible without using open science in one way or another.

For the large pool of microscopists for whom biological discovery is the key driver, the goal is not to apply the method with the best resolution on paper. Rather, the aim is to find or develop the most suitable technique, or combination of techniques, that work within the constraints of a specific biological question. These researchers benefit from the modular nature of open microscopy when they rapidly test, prototype, and tailor different microscopy approaches for their specific system, and often combine multiple advanced microscopy techniques in a way that simply would not be feasible in either commercial or traditional home-built systems. This allows researchers to use the best microscopy tool for their project, instead of being limited by what is available in their local facility or needing to embark on multi-year fundraising efforts. Above all, open microscopy opens up the black box of technology-driven device development and makes it more accessible to those who use it. Creative thinking that emerges from this can produce new methods and a fully interdisciplinary approach to research.

Despite the progress and promise of open microscopy, long term support for open microscopy projects, accessibility of more complex designs for non-specialists, and incomplete availability of documentation and designs remain outstanding problems in the field. To solve these challenges, we outlined several possible solutions. Sharing microscope technology using openly accessible and modifiable documentation provides the opportunity to create communities that can ensure the long-term stability and continuing development that end-users are hoping for when investing time and resources in open microscopy. Outsourcing or commercialization of open microscopy hardware could largely solve the sustainability problem and increase accessibility for the end-user.

At its best open microscopy empowers scientific curiosity, creativity, and collaboration. For this reason alone, it is worth investing time and money into its bright future.

## Additional resources

A list of hardware and software projects, repositories and additional resources can be found on [https://github.com/HohlbeinLab/OpenMicroscopy](https://github.com/HohlbeinLab/OpenMicroscopy). The authors welcome contributions to make the list comprehensive and keep the list up to date.

## Author contributions

J.H. and K.P. initiated the manuscript. All authors provided sections and contributed to the final version of the text.

## Acknowledgements

The authors thank their colleagues and SciTwitter for inspiration and discussions. We thank Lothar Schermelleh for discussions and suggestions; Nikita Vladimirov for his kind help with the MesoSPIM figure.  R.H. acknowledges support by the Deutsche Forschungsgemeinschaft under Germany's Excellence Strategy - EXC2068 - Cluster of Excellence Physics of Life of TU Dresden. We thank Dimitrios Tsikritsis for providing valuable feedback to the manuscript.

**Figures**

*Figure 1*

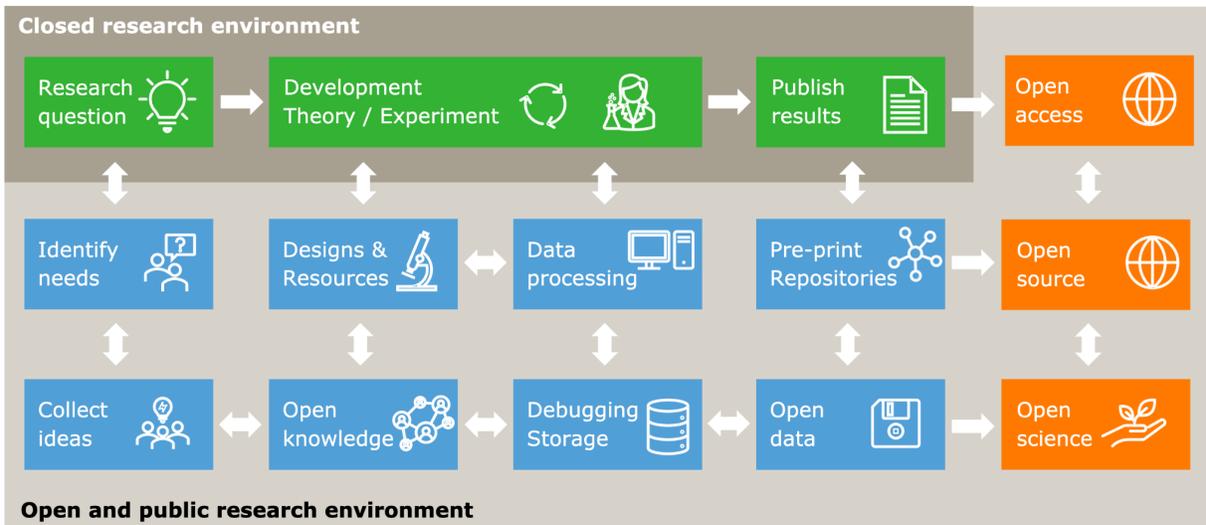

**Figure 1.** Open science can accelerate scientific progress. Instead of operating in a closed environment, representing, for example, individual labs, research units or scientific areas, free flow of information and exchange of ideas, resources and data enables the efficient pooling of resources.

*Figure 2*

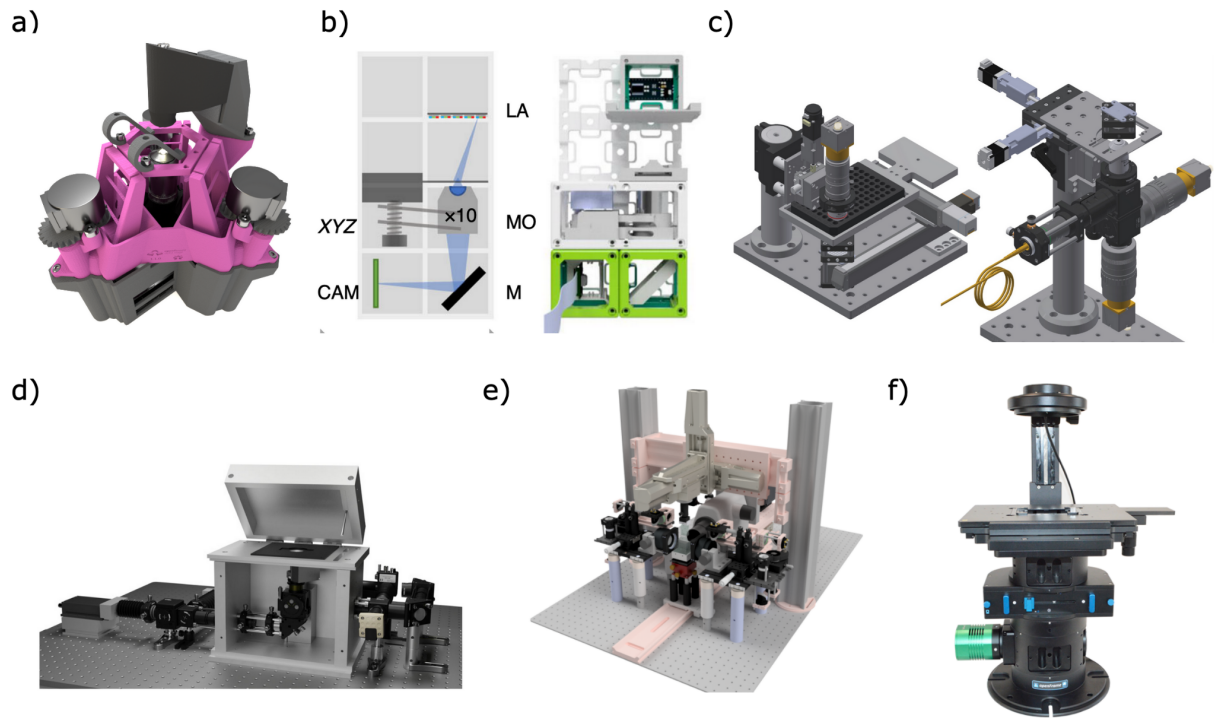

**Figure 2.** Overview on various open-microscopy projects. **a)** OpenFlexure[3], **b)** "You see, too" (UC2[4]), **c)** Simplifying Quantitative Imaging Platform Development and Deployment (SQUID[27]), **d)** single-molecule Förster resonance energy transfer box (smFRETbox[29]), **e)** MesoSPIM[20], **f)** openFrame[34], an open microcopy framework that is commercially available. Permissions: images a), b), d) reproduced under CC-BY 4.0, c), e) under CC-BY-NC-ND 4.0 and f) with permission from Cairn-Research).